\documentclass{emulateapj}

\newcommand{\etal}{et~al.~}

\newcommand{\chisqr}{$\chi^2$}

\newcommand{\kms}{\ifmmode\,{\rm km}\,{\rm s}^{-1}\else km$\,$s$^{-1}$\fi} 
 
\def \spose#1{\hbox to 0pt{#1\hss}}                                   
\def \lta{\mathrel{\spose{\lower 3pt\hbox{$\sim$}}                  
     \raise 2.0pt\hbox{$<$}}}                                                 
\def \gta{\mathrel{\spose{\lower 3pt\hbox{$\sim$}}                   
     \raise 2.0pt\hbox{$>$}}}

\def\Equ#1{Eq.~(\ref{eq:#1})}                
\def\Fig#1{Fig.~\ref{fig:#1}}

\def\be{\begin{equation}}                                         
\def\ee{\end{equation}}                                         
                                                     
\def\ifm#1{\relax\ifmmode#1\else$\mathsurround=0pt #1$\fi}  
\def\kms{\ifmmode\,{\rm km}\,{\rm s}^{-1}\else km$\,$s$^{-1}$\fi} 
        
\def\kpc{\,{\rm kpc}}

\def\msol{M_{\odot}}

\def\MBH {M_{\small BH}}

\def\hmsun{h^{-1}\msol}  
\def\hMsun{\hmsun}  
\def\ltsima{$\; \buildrel < \over \sim \;$}                    
\def\lsim{\lower.5ex\hbox{\ltsima}}  
\def\gtsima{$\; \buildrel > \over \sim \;$}                            
\def\gsim{\lower.5ex\hbox{\gtsima}} 
 
\def\Vc{V_{\rm c}}  
  
\def\Vr{V_{\rm rot}}  
\def\Rm{R_{\rm max}}

\def\cs{\sigma_0} 
\def\C28{\rm C_{28}} 

\def\pmb#1{\setbox0=\hbox{#1}%
\kern-.025em\copy0\kern-\wd0 \kern.05em\copy0\kern-\wd0 
\kern-.025em\raise.0433em\box0} 
 
\def \ion#1#2{#1{\footnotesize{#2}}\relax}  
\def \ha {H$\alpha$\ }  
\def \hi {\ion{H}{I} }

\def \littlemm{\ifmmode{\scriptscriptstyle m } 
     \else{\hbox{$\scriptscriptstyle m $ }}\fi}  
\def \topemm{\raise .9ex \hbox{\littlemm}}  
\def \magpoint{\hbox to 2pt{}\rlap{\hskip -.5ex 
     \topemm}.\hbox to 2pt{}}  

\usepackage{apjfonts}
\usepackage{lscape}

\def \etal {et~al.~}
\def \chisq  {\ifmmode  \chi^2   \else  $\chi^2$  \fi}  
\def \chisqr {\ifmmode \chi^2_{\rm r} \else $\chi^2_{\rm r}$ \fi}
\def \spose#1{\hbox  to 0pt{#1\hss}}  
\def \lta{\mathrel{\spose{\lower 3pt\hbox{$\sim$}}\raise  2.0pt\hbox{$<$}}}
\def \gta{\mathrel{\spose{\lower  3pt\hbox{$\sim$}}\raise 2.0pt\hbox{$>$}}}
\def \ion#1#2{#1{\footnotesize{#2}}\relax} 
\def \ha  {\ifmmode H\alpha \else H$\alpha $ \fi} 
\def \hi {\ion{H}{I}}

\def \kms {\ifmmode  \,\rm km\,s^{-1} \else $\,\rm km\,s^{-1}  $ \fi }
\def \kpc {\ifmmode  {\rm kpc}  \else ${\rm  kpc}$ \fi  }  
\def \Msun {\ifmmode M_{\odot} \else $M_{\odot}$ \fi} 
\def \hMsun {\ifmmode h^{-1}\,\rm M_{\odot} \else $h^{-1}\,\rm M_{\odot}$ \fi}
\def \hhMsun {\ifmmode h^{-2}\,\rm M_{\odot}\else $h^{-2}\,\rm M_{\odot}$ \fi}
\def \Lsun {\ifmmode L_{\odot} \else $L_{\odot}$ \fi} 
\def \hhLsun {\ifmmode h^{-2}\,\rm L_{\odot} \else $h^{-2}\,\rm L_{\odot}$ \fi}
\newcommand{\magarc}{\ifmmode {{{{\rm mag}~{\rm arcsec}}^{-2}}}
             \else {{{mag}$~${arcsec}$^{-2}$}}
             \fi}

\def \LCDM {\ifmmode \Lambda{\rm CDM} \else $\Lambda{\rm CDM}$ \fi}
\def \sig8 {\ifmmode \sigma_8 \else $\sigma_8$ \fi} 
\def \OmegaM {\ifmmode \Omega_{\rm m} \else $\Omega_{\rm m}$ \fi} 
\def \OmegaL {\ifmmode \Omega_{\rm \Lambda} \else $\Omega_{\rm \Lambda}$\fi} 
\def \Deltavir {\ifmmode \Delta_{\rm vir} \else $\Delta_{\rm vir}$ \fi}

\def \aac {\ifmmode \alpha_{\rm AC} \else $\alpha_{\rm AC}$ \fi} 
\def \rs {\ifmmode r_{\rm s} \else $r_{\rm s}$ \fi} 
\def \rrm2 {\ifmmode r_{-2} \else $r_{-2}$ \fi} 
\def \ccm2 {\ifmmode c_{-2} \else$c_{-2}$ \fi} 
\def \cvir {\ifmmode c_{\rm vir} \else $c_{\rm vir}$ \fi} 
\def \cbar {\ifmmode \overline{c} \else $\overline{c}$ \fi}

\def \R200 {\ifmmode R_{200} \else $R_{200}$ \fi} 
\def \Rvir {\ifmmode R_{\rm vir} \else $R_{\rm vir}$ \fi}

\def \v200 {\ifmmode V_{200} \else $V_{200}$ \fi} 
\def \Vvir {\ifmmode V_{\rm  vir} \else  $V_{\rm vir}$  \fi} 
\def  \Vhalo  {\ifmmode V_{\rm halo} \else $V_{\rm halo}$ \fi}

\def \M200 {\ifmmode M_{200} \else $M_{200}$ \fi} 
\def \Mvir {\ifmmode M_{\rm  vir} \else $M_{\rm  vir}$ \fi}  
\def \Mshell  {\ifmmode M_{\rm shell} \else $M_{\rm shell}$ \fi}

\def \Jvir {\ifmmode J_{\rm vir} \else $J_{\rm vir}$ \fi} 
\def \Jshell {\ifmmode J_{\rm shell} \else $J_{\rm shell}$ \fi}

\def \lamgal {\ifmmode \lambda_{\rm gal}  \else $\lambda_{\rm gal}$ \fi} 
\def \lam {\ifmmode \lambda  \else $\lambda$ \fi} 
\def \lamp {\ifmmode \lambda^{\prime} \else $\lambda^{\prime}$  \fi} 
\def \lambar {\ifmmode \bar{\lambda}  \else  $\bar{\lambda}$  \fi}  
\def \lampbar  {\ifmmode \bar{\lambda^{\prime}} \else $\bar{\lambda^{\prime}}$\fi} 
\def \siglam {\ifmmode \sigma_{\lambda} \else $\sigma_{\lambda}$ \fi} 
\def \siglamp {\ifmmode \sigma_{\lambda^{\prime}} \else $\sigma_{\lambda^{\prime}}$\fi}  

\def \qd {\ifmmode q_{\rm d} \else $q_{\rm d}$ \fi} 
\def \qh {\ifmmode q_{\rm h} \else $q_{\rm h}$ \fi}

\def \dd {\ifmmode {\rm d} \else ${\rm d}$ \fi} 

\def \alpham {\ifmmode \alpha_{\rm m} \else $\alpha_{\rm m}$ \fi} 
\def \fbar {\ifmmode f_{\rm bar} \else $f_{\rm bar}$ \fi} 
\def \RI {\ifmmode R_{I} \else $R_{I}$ \fi} 
\def \Rd {\ifmmode R_{\rm d} \else $R_{\rm d}$ \fi} 
\def \Rs {\ifmmode R_{*} \else $R_{*}$ \fi}  
\def \Rd {\ifmmode R_{\rm d} \else $R_{\rm d}$ \fi}  
\def \Rcool  {\ifmmode R_{\rm  cool} \else $R_{\rm cool}$ \fi} 
\def \RIII {\ifmmode  3.2\Rs \else $3.2\Rs$ \fi} 
\def \RII {\ifmmode 2.2\Rs \else $2.2\Rs$  \fi} 
\def \Reff {\ifmmode R_{\rm eff} \else $R_{\rm  eff}$ \fi} 
\def \rb {\ifmmode r_{\rm b}  \else $r_{\rm b}$ \fi}
\def \ri {\ifmmode r_{\rm i}  \else $r_{\rm i}$ \fi}
\def \rf {\ifmmode r_{\rm f}  \else $r_{\rm f}$ \fi}

\def \Sigmacrit {\ifmmode \Sigma_{\rm  crit} \else $\Sigma_{\rm crit}$\fi} 
\def \Sig0 {\ifmmode \Sigma_{0} \else $\Sigma_{0}$ \fi}

\def \muI {\ifmmode \mu_{0,I} \else $\mu_{0,I}$ \fi}

\def \mgal    {\ifmmode m_{\rm gal}    \else $m_{\rm gal}$ \fi} 
\def \mgalo    {\ifmmode m_{\rm gal,0}    \else $m_{\rm gal,0}$ \fi} 
\def \md    {\ifmmode m_{\rm d}    \else $m_{\rm d}$ \fi} 
\def \ms    {\ifmmode m_{\rm s}    \else $m_{\rm s}$ \fi}   
\def \Md    {\ifmmode M_{\rm d}    \else $M_{\rm d}$ \fi} 
\def \Ms    {\ifmmode M_{*}    \else $M_{*}$ \fi} 
\def \Mb    {\ifmmode M_{\rm b}    \else $M_{\rm b}$ \fi} 
\def \Mf    {\ifmmode M_{\rm f}    \else $M_{\rm f}$ \fi} 
\def \Mi    {\ifmmode M_{\rm i}    \else $M_{\rm i}$ \fi} 
\def \Mgal  {\ifmmode M_{\rm gal}  \else $M_{\rm gal}$ \fi}
\def \Mstar {\ifmmode M_{\rm star} \else $M_{\rm star}$ \fi}
\def \Mdisk {\ifmmode M_{\rm disk} \else $M_{\rm disk}$ \fi}
\def \msbar {\ifmmode \bar{m}_{\rm s} \else $\bar{m}_{\rm s}$ \fi}  
\def \mdbar {\ifmmode {\overline{m}}_{\rm d} \else ${\overline{m}}_{\rm d}$ \fi} 

\def \Jd {\ifmmode J_{\rm d} \else $J_{\rm d}$ \fi} 
\def \Jb {\ifmmode J_{\rm b} \else $J_{\rm b}$ \fi}  
\def \Jgal {\ifmmode J_{\rm gal} \else $J_{\rm gal}$ \fi}  
\def \fb {\ifmmode f_{\rm b} \else $f_{\rm b}$ \fi}
\def \jd   {\ifmmode j_{\rm  d}  \else  $j_{\rm  d}$ \fi}  
\def \jgal   {\ifmmode j_{\rm  gal}  \else  $j_{\rm  gal}$ \fi}  
\def \jdmd {\ifmmode \frac{j_{\rm  d}}{m_{\rm d}} \else $\frac{j_{\rm d}}{m_{\rm d}}$ \fi} 
\def \fc {\ifmmode f_{\rm c} \else $f_{\rm c}$ \fi} 
\def \fx {\ifmmode f_{\rm x} \else $f_{\rm x}$ \fi} 
\def \fj {\ifmmode f_{\rm j} \else $f_{\rm j}$ \fi} 
\def \ft {\ifmmode f_{\rm t} \else $f_{\rm t}$ \fi} 
\def \fM {\ifmmode f_{\rm M} \else $f_{\rm M}$ \fi}
\def \fV {\ifmmode f_{\rm V} \else $f_{\rm V}$ \fi}
\def \fmd {\ifmmode m_{\rm d} \else $m_{\rm d}$ \fi} 
\def \fjd {\ifmmode j_{\rm d} \else $j_{\rm d}$ \fi} 
\def \fmb {\ifmmode m_{\rm b} \else $m_{\rm b}$ \fi} 
\def \fjb {\ifmmode j_{\rm b} \else $j_{\rm b}$ \fi} 
\def \fmgal {\ifmmode m_{\rm gal} \else $m_{\rm gal}$ \fi} 
\def \fjgal {\ifmmode j_{\rm gal} \else $j_{\rm gal}$ \fi} 
\def \flost {\ifmmode f_{\rm lost} \else $f_{\rm lost}$ \fi} 

\def \Vc    {\ifmmode  V_{\rm c}    \else $V_{\rm c}$    \fi} 
\def \Vd    {\ifmmode  V_{\rm d}    \else $V_{\rm d}$    \fi} 
\def \Vb    {\ifmmode  V_{\rm b}    \else $V_{\rm b}$    \fi} 
\def \VDM   {\ifmmode  V_{\rm DM}   \else $V_{\rm DM}$    \fi} 
\def \Vcd   {\ifmmode  V_{\rm c,d}  \else $V_{\rm c,d}$  \fi} 
\def \Vcb   {\ifmmode  V_{\rm c,b}  \else $V_{\rm c,b}$  \fi} 
\def \VcDM  {\ifmmode  V_{\rm c,DM} \else $V_{\rm c,DM}$ \fi} 
\def \Vcool {\ifmmode  V_{\rm cool} \else $V_{\rm cool}$ \fi} 
\def \Vcirc {\ifmmode  V_{\rm circ} \else $V_{\rm circ}$ \fi} 
\def \VIII  {\ifmmode  V_{3.2}      \else $V_{3.2}$      \fi} 
\def \VII   {\ifmmode  V_{2.2}      \else $V_{2.2}$      \fi}
\def \Vobs  {\ifmmode  V_{\rm obs}  \else $V_{\rm obs}$  \fi} 
\def \Vdisk {\ifmmode  V_{\rm disk} \else $V_{\rm disk}$ \fi} 
\def \Vmax  {\ifmmode  V_{\rm  max} \else $V_{\rm max}$  \fi} 
\def \Vtot  {\ifmmode  V_{\rm tot}  \else $V_{\rm tot}$  \fi} 
\def \Vrot  {\ifmmode  V_{\rm rot}  \else $V_{\rm rot}$  \fi} 
\def \Vflat {\ifmmode  V_{\rm  flat}\else $V_{\rm flat}$ \fi}
\def \Vmaxobs{\ifmmode V_{\rm max}^{\rm obs}\else $V_{\rm max}^{\rm obs}$\fi}  

\def \vdt {\ifmmode  \Vd/\VII \else  $\Vd/\VII$ \fi}  
\def \dvr {\ifmmode \partial \log  V_{2.2} / \partial \log R_{\rm d} 
     \else  $\partial \log V_{2.2} / \partial \log R_{\rm d}$ \fi} 
\def \Dvr  {\ifmmode \Delta \log V / \Delta \log R \else $\Delta \log
  V / \Delta \log R$ \fi}
\def \DVL  {\ifmmode \Delta \log V | L_I \else $\Delta \log
  V | L_I$ \fi}
\def \DRL  {\ifmmode \Delta \log\Rs | L_I \else $\Delta \log
  \Rs | L_I$ \fi}
\def \DRV  {\ifmmode \Delta \log\Rs | V \else $\Delta \log
  \Rs | V$ \fi}
\def \DLV  {\ifmmode \Delta \log L_I | V \else $\Delta \log
  L_I | V$ \fi}
\def \DVR  {\ifmmode \Delta \log V | \Rs \else $\Delta \log
  V | \Rs$ \fi}
\def \DLR  {\ifmmode \Delta \log L_I | \Rs \else $\Delta \log
  L_I | \Rs$ \fi}

\def \Ups {\ifmmode \Upsilon  \else $\Upsilon$ \fi} \def \Yd {\ifmmode
\Upsilon_{\rm  d} \else  $\Upsilon_{\rm d}$  \fi} \def  \YdR {\ifmmode
\Upsilon_{\rm d}^R \else $\Upsilon_{\rm  d}^R$ \fi} \def \YB {\ifmmode
\Upsilon_B \else $\Upsilon_B$ \fi} \def \Yr {\ifmmode \Upsilon_r \else
$\Upsilon_r$  \fi} \def  \YI {\ifmmode  \Upsilon_I  \else $\Upsilon_I$
\fi} \def  \YH {\ifmmode \Upsilon_H  \else $\Upsilon_H$ \fi}  \def \YK
{\ifmmode \Upsilon_{K'} \else  $\Upsilon_{K'}$ \fi} \def \LI {\ifmmode
L_I \else $L_I$ \fi}

\def  \alphacrit   {\ifmmode  \alpha_{\rm  crit}   \else  $\alpha_{\rm
crit}$\fi}

\def \DeltaIMF {\ifmmode \Delta_{\rm IMF} \else $\Delta_{\rm IMF}$ \fi}

\slugcomment{Accepted for publication in ApJL} 

\shorttitle{The Bulge-Halo Connection in Galaxies} 
\shortauthors{Courteau \etal 2006}

\begin{document}


\title{The Bulge-Halo Connection in Galaxies: A Physical Interpretation 
       of the $\Vc-\cs$ Relation}

\author{St\'ephane Courteau, Michael McDonald, Lawrence M. Widrow}    
\affil{Department of Physics, Engineering Physics \& Astronomy, Queen's
  University, Kingston, ON \ K7L 3N6, Canada}
\and

\author{Jon Holtzman}  
\affil{Department of Astronomy, New Mexico State University, Las Cruces, 
       NM \ 88003, USA}
\email{courteau,mmcdonald,widrow@astro.queensu.ca; holtz @nmsu.edu} 

\begin{abstract}
  
We explore the dependence of the ratio of a galaxy's maximum circular 
velocity, $\Vc$, to its central velocity dispersion, $\cs$, on morphology, 
or equivalently total light concentration.
Such a dependence is expected if light traces the mass.  
Over the full range of galaxy types, masses and brightnesses, and 
assuming that the gas velocity traces the circular velocity, we find 
that galaxies obey the relation $\log(\Vc/\cs)= 0.63 - 0.11 C_{28}$
where $C_{28}=5\log(r_{80}/r_{20})$ and the radii are measured at 
80\% and 20\% of the total light.  Massive galaxies scatter 
about the $\Vc = \sqrt{2} \cs$ line for isothermal stellar 
systems.  
For pure disks, $\C28\sim 2.8$ 
and $\Vc \simeq 2\cs$. 
Self-consistent equilibrium galaxy models from Widrow \& Dubinski (2005) 
constrained to match the size-luminosity and velocity-luminosity relations
of disk galaxies fail to match the observed $\Vc/\cs$ distribution.
Furthermore, the matching of dynamical models for $\Vc(r)/\sigma(r)$ 
with observations of dwarf and elliptical galaxies suffers from limited 
radial coverage and relatively large error bars; for dwarf systems, 
however, kinematical measurements at the galaxy center and optical 
edge suggest $\Vc(\Rm) \gta 2 \cs$ (in contrast with past assumptions 
that $\Vc(\Rm) = \sqrt{2} \cs$ for dwarfs.)  The $\Vc-\cs-C_{28}$ relation 
has direct implications for galaxy formation and dynamical models, galaxy 
scaling relations, the mass function of galaxies, and the links between 
the formation and evolution processes of a galaxy's central 
massive object, bulge, and dark matter halo.  
 
\end{abstract}

\keywords{galaxies: bulge --- 
	  galaxies: elliptical and lenticular ---
          galaxies: fundamental parameters  ---
          galaxies: kinematics and dynamics ---
          galaxies: spirals ---
	  dark matter}



\section{Introduction}
\label{sec:intro}

Much excitement has followed the discovery of a connection between
a galaxy's central black hole mass and the bulge velocity dispersion 
(Ferrarese \& Merritt 2000; Gebhardt \etal 2000) $\MBH-\cs$.  When 
coupled with the relation between the  
deprojected circular velocity, $\Vc(r)= [GM(r)/r]^{1/2}$, where $M(r)$ 
is the total mass within $r$ of the center, and the one-dimensional 
line-of-sight (projected) central velocity dispersion, $\cs$, a 
relation between supermassive black hole (SBH) and total galaxy 
masses can be established (Ferrarese 2002 [F02]; Baes \etal 2003 [B03]; 
Pizzella \etal 2005 [P05]).  Such a connection
supports the notion of regulated formation mechanisms and co-evolution 
for the smallest and largest structures in galaxies (e.g. F02; Ho 2004;
Ferrarese \etal 2006 [F06]). 

Previous derivations of the $\Vc-\cs$ relation for spiral and elliptical 
galaxies for which dynamical measurements were either measured and/or 
inferred (Whitmore \etal 1979 [WKS79]; F02; B03; P05), have considered 
a simple relation of the form $\log\Vc = a~\log\cs + b$.  
For spiral galaxies, $\Vc$ is measured at some fiducial radius 
in their outer parts; for ellipticals, $\Vc$ is inferred via dynamical 
modeling of the absorption line features and surface brightness profiles.  
The dynamical modeling of elliptical systems is, however, intrinsically 
challenging and the resulting dynamical estimates remain significantly 
uncertain. For a heterogeneous sample (described below) of 40 high 
surface brightness (HSB) spiral and 24 elliptical galaxies, P05 
reported the one-dimensional correlation:


\begin{equation}
\label{eq:P05}
\log \Vc = (0.74 \pm 0.07) \log\cs + (0.80\pm 0.15).
\end{equation}


The realisation that elliptical and spiral galaxies appear 
to obey the same $\Vc-\cs$ relation\footnote{Not 
to be confused with the $V_\circ/\cs$ diagnostic of
Kormendy \& Illingworth (1982; also Binney \& Tremaine Fig. 4-6) 
for elliptical galaxies which uses different 
velocity measurements; Kormendy \& Illingworth measure $V_\circ$ 
in the bulge (e.g. at the bulge effective radius) whereas our
$\Vc$ is a measurement near the peak of the rotation curve.} 
prompted P05 to suggest that the central velocity dispersion 
of a galaxy is independent of morphological type for a given 
dark matter halo.  Assuming that mass traces light, the opposite 
is, however, expected on first dynamical principles. 

We can write the Jeans equation for spherical, or nearly
spherical, self-gravitating systems (e.g. WKS79; 
Binney \& Tremaine 1987, Eq.~4-55; see also Dekel \etal 2005):

\begin{equation}
\label{eq:BT87} 
\Vc^2(r) = [\alpha(r) + \gamma(r) - 2\beta(r)]\sigma_r^2(r)
\end{equation}
where $\alpha$ is the logarithmic derivative of the stellar
density profile $\nu$, $\gamma \equiv -d\ln\sigma_r^2/d\ln{r}$, 
and $\beta \equiv 1 - \sigma_\theta^2/\sigma_r^2$ is the 
anisotropy parameter.
We can rewrite \Equ{BT87} as:

\begin{equation}
\label{eq:Jeans}
\frac{V_{c}}{\sigma_{r,in}} = \frac{V_{c}}{V_{c,in}} [\alpha_{in}
  + \gamma_{in} - 2\beta_{in}]^{1/2}
\end{equation}
where the index ``in'' refers to quantities measured in 
the inner parts of the galaxy, by contract with $\Vc$ which is
measured in the outer parts.

The ratio $\Vc/\cs \sim \Vc/\sigma_{r,in}$ depends strongly on 
the shape of the underlying rotation curve, through $\Vc/V_{c,in}$, 
which is itself directly related to morphology or concentration (e.g. 
Kent 1987).  And while a dependence of $\alpha$, $\beta$, and $\gamma$ 
on concentration is also expected on first principles, 
a full derivation is thwarted by the non-trivial covariances between 
these terms and the limited data to decompose $\Vc/\cs$ in terms of 
the Jeans equation parameters.  Still, we verify empirically in 
this {\it Letter}, in accord with WKS79 but in contrast with P05, 
the dependence of the $\Vc/\cs$ ratio on concentration. 

The simultaneous fitting of $\Vc(r)$, $\sigma(r)$, and the light 
profile, $L(r)$, to fully constrain galaxy dynamics has 
thus far been attempted mostly for dwarf galaxies which are 
dark matter dominated at nearly all radii and where concerns 
about stellar populations are lessened (e.g. Dutton \etal 2005).  
Such dynamical mapping of dwarf galaxies was explored by 
many as a potential route to solving the so-called ``satellite 
over-production'' problem
in $\Lambda$CDM structure formation models since the 
same physics to predict galaxy satellite distributions and internal 
dynamics is at play (e.g. Moore \etal 1999; Stoehr \etal 2002; 
Kazantzidis \etal 2004; Mashchenko, Sills, \& Couchman 2006).  
These models involve significant simplifications and solutions 
are not unique; a major limitation is the degree of velocity dispersion 
anisotropy which remains poorly constrained with current observations; 
there indeed exists a well-known degeneracy between velocity 
anisotropy and mass distribution given only rotation and velocity 
dispersion data (e.g. Gerhard \etal 1998), as well as
significant degeneracy in the $\Vc - \Rm$ values of the host 
halo that reproduce observed $\sigma(r)$ and surface 
brightness profiles (Bullock 2006; priv. comm.)  Furthermore,
dynamical studies of dwarf galaxies have often assumed 
$\Vc \simeq \sqrt{2} \cs$, the expectation for an isothermal 
sphere (Binney \& Tremaine 1987; Eq. 4-127b).  
We show in this {\it Letter} that this assumption is unlikely.

While the mass modeling of dwarf, and more massive, galaxies in 
the context of $\Lambda$CDM cosmogony still falls short of providing 
a clear mapping of $\Vc(r)/\sigma(r)$ with galaxy types, progress 
can still be achieved by considering specific parameters such as 
the maximum circular velocity, $\Vc$, the central velocity dispersion,
$\cs$, and their relation across the full range of galaxy types and 
masses.  

Below we use existing data bases to demonstrate
the dependence of $\Vc/\cs$ on the galaxy light concentration.  
We attempt to bolster this result in \S4 using self-consistent 
dynamical models and conclude with a discussion of the significance 
of this result for models of galaxy dynamics and formation.


\section{The Data}
\label{sec:data}

We have compiled data from the literature which includes
measurements or modeling of $\Vc$ and $\cs$ for spiral, lenticular, 
and elliptical galaxies.
We have first built upon the compilation of $\Vc$ and $\cs$ by 
P05 from the heterogeneous data bases of F02 and B03. P05 added data 
of their own for a total of 40 high surface brightness (HSB) galaxies 
of types S0/a to Scd and 8 low surface brightness (LSB) galaxies 
of types Sa to Sc.  These 48 spiral galaxies 
were selected to have well-defined flat rotation curves.  
Following F02, P05's galaxy list includes 20 elliptical galaxies 
whose $\Vc$ was inferred by non-parametric dynamical modeling 
of the absorption line features and surface brightness profiles 
(Kronawitter \etal 2000 [K00]; Gerhard \etal 2001; see also 
Gerhard \etal 1998 and Cappellari \etal 2006).  To these, P05 
added 5 elliptical galaxies with inner gaseous disks whose 
$\Vc$ could be measured via \hi\ linewidths.  
We have expanded P05's galaxy compilation with the independent 
sample of 66 spiral galaxies with measurements
of $\Vc$ and $\cs$ by Prugniel, Maubon, \& Simien (2001 [PMS01]). 
These authors measured central stellar velocity dispersions from 
absorption features of galaxy spectra and maximum circular 
velocities were derived from \hi\ line widths\footnote{Provided 
suitable measurements, \hi\ line widths and optical line 
velocities trace the same dynamics (Courteau 1997).}.  
Bedregal \etal (2006; B06)
also compiled heterogeneous kinematic data for 51 S0 galaxies
that we shall use; the stellar kinematics to estimate $\Vc$ were 
all corrected for asymmetric drift following Neistein \etal 1999.  
Finally, in order to map the lower range of galaxy masses and 
brightnesses, we have also considered the dynamical measurements 
for 8 Local Group dwarf irregular (dI) galaxies with $\Vc \gsim 10$ 
\kms by Mateo (1998) and Woo \etal (2006).


Our heterogeneous sample includes 154 HSB/LSB spiral, 54 lenticular, 
24 elliptical, and 11 dwarf irregular galaxies with measured $\cs$ 
and either measured or inferred $\Vc$.  Our compilation, including 
new concentrations presented below, is available at 
www.astro.queensu.ca/$\sim$courteau/data/VSigmaC28.txt.

Based on an overlap of 10 galaxies between F02 and PMS01, we estimate 
a systematic difference of 24 \kms for $\Vc$ and 9 \kms for $\cs$
over the full range of $\Vc$ and $\cs$.  
There is no overlap between the Southern hemisphere galaxies compiled 
by B03 and the samples of PMS01, F02, or B06.  Our sub-samples are all 
collections of heterogeneous data and it is not clear, due to the
small sample sizes, whether any bias exists among the data sets.  
We make the assumption that our compiled data for $\Vc$ and $\cs$ 
are uniformly calibrated and can be inter-compared.
As supporting evidence we note that our main conclusions hold 
whether we analyze the individual samples or the full data collection.  
One must also keep in mind that $\cs$'s 
in spiral galaxies may be polluted by disk stars and thus represent 
an upper limit.  We also make the assumption that the rotational velocity
of the gas, $\Vr$, is an accurate tracer of the circular velocity, i.e. 
$\Vc \equiv \Vr$ for spirals. 

\begin{figure*}
\begin{center}
\includegraphics[width=7.0in]{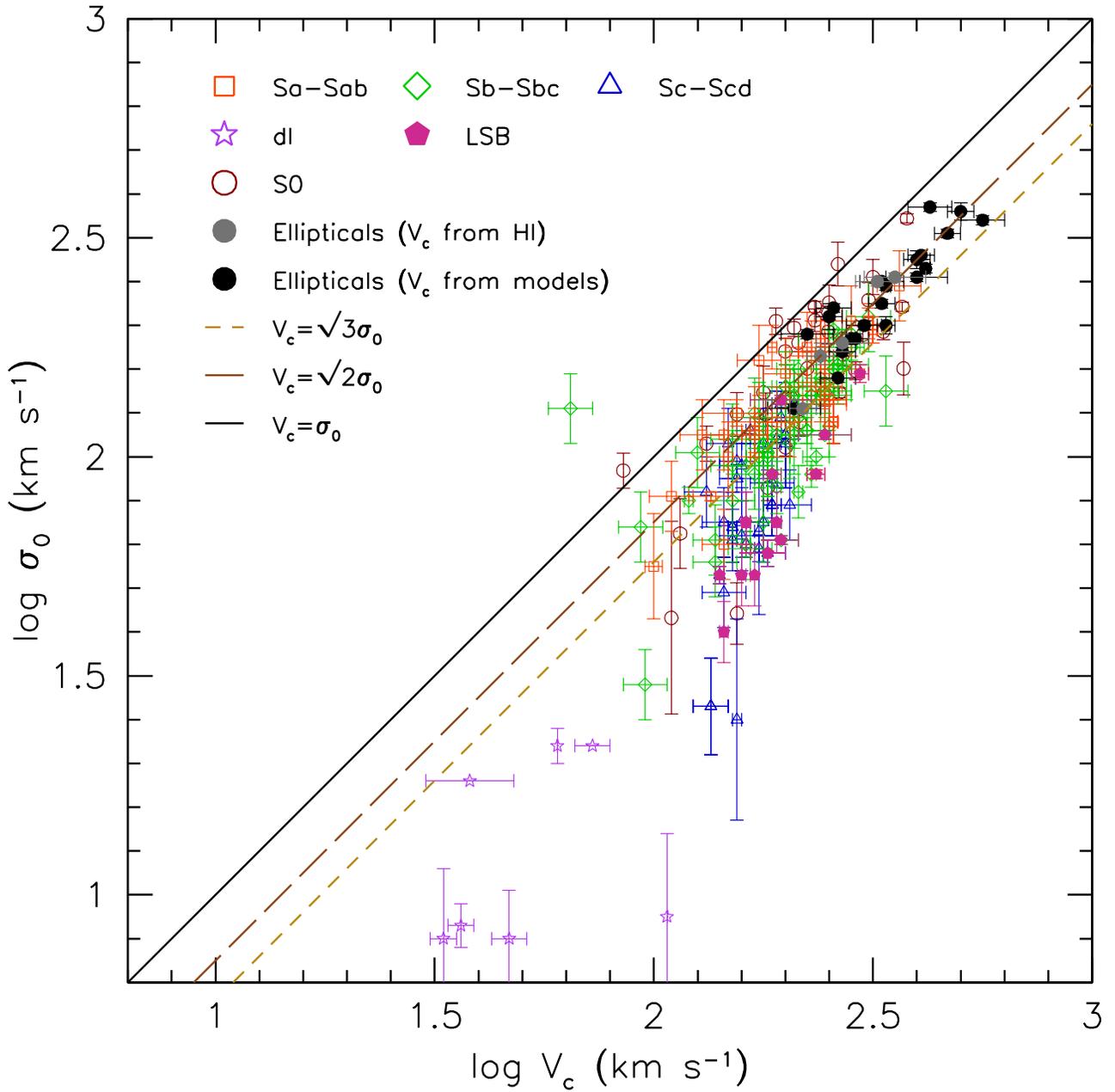}
\caption{Maximum circular velocity, $\Vc$, versus central velocity 
  dispersion, $\cs$, for 154 spiral, 24 elliptical, 54 lenticular, 
  and 11 dwarf irregular galaxies.
  It is assumed that gas velocities in spiral galaxies trace 
  circular motion.  The spiral systems show a clear 
  dependence on morphological type such that, for a given dark 
  matter halo ($\Vc$), earlier-type galaxies have a higher $\cs$.  
  Massive galaxies scatter about the $\Vc=\sqrt{2}\cs$ line for 
  isothermal stellar systems, with significant departures from 
  the isothermal expectation for systems with $\Vc \lta 200$ \kms.}
\label{fig:vsigma}
\end{center}
\end{figure*}

\Fig{vsigma} shows the distribution of $\Vc$ and $\cs$ for our full
sample.  A well defined, though broad, $\Vc-\cs$ relation seems to hold 
for galaxies with $\cs \gta 80 \kms$ and $\Vc \gta 200 \kms$.
For smaller galaxies, rotational and dispersion estimates are less 
certain due to relatively more prominent gas turbulence, velocity 
anisotropy, and measurement errors.

The morphological dependence of the $\Vc-\cs$ 
relation for spiral galaxies is obvious.  For a given $\Vc$, or
total luminosity, early-type spirals have a higher $\cs$ than 
later types (WKS79).  For the sample of 21 ellipticals 
reported in P05, K00 find that the anisotropy parameter 
$\beta \lta 0.3$ at $R_e/2$ and most of their galaxies exhibit 
near-isotropy near the center.  This sample is dominated by E0-E1 
ellipticals that scatter about the isothermal $\Vc \simeq \sqrt{2} \cs$ 
line.  The possible dependence of $\Vc-\cs$ on morphology for elliptical 
galaxies is discussed below. 

Note how LSB galaxies delineate the envelope of spiral
galaxies in the $\Vc-\cs$ distribution on account 
of their relatively small bulges, or low concentrations. P05 
viewed the distributions of LSB and HSB spirals as two distinct 
$\Vc-\cs$ relations but fit HSB spiral and elliptical galaxies 
with a common $\Vc-\cs$ relation (Eq.~1); this 
statement, however, overlooks the expected $\Vc-\cs$ dependence 
on galaxy concentration for all galaxy types. 
We quantify this relation in \S{3}.

The comparison of $\Vc-\cs$ for ellipticals with \hi\ 
disks demands care as the inner disks within flattened 
ellipticals 
are embedded in a complex triaxial potential, unlike pure spiral disks 
which may revolve in a more spherical halo and for which the observed 
radial velocity is assumed to trace the mass. 
For the E1-2 (Sy1) galaxy NGC 4278 as reported in P05, 
$\Vc$(\hi) $= 326 \pm 40$ while $\Vc$ ({\small model}) $= 416 \pm 13$
making for a rather ambiguous interpretation of $\Vc$.  Indeed, the
measurement and modelling of $\sigma(r)$ and $\Vc(r)$ for 
E galaxies, much like dwarf systems (\S{1}), is complex and 
suffers from limited radial coverage and relatively large error bars.  
Large scale dynamical modeling of E galaxies has only been 
attempted for a few bright systems so far (e.g. K00).  
Repeat observations and measurements of new systems by independent
teams are direly needed. 


\section{Light Concentration}
\label{sec:dynamics}

To assess the dependence of $\Vc/\cs$ on galaxy structure for all 
galaxy types, we use the total light concentration: 

\begin{equation}  
\label{eq:c28}  
\C28=5\log(r_{80}/r_{20}),
\end{equation}  
where $r_{20}$ and $r_{80}$ are the radii within which 20\% and 80\% 
of the total light is contained (Kent 1987).

We compute concentrations from Sloan Digital Sky Survey 
(York \etal 2000, [SDSS]) $i$-band images for 32 spiral, 
12 elliptical, and 33 lenticular galaxies in our sample.  
We have verified that the concentrations extracted from SDSS 
multi-band galaxy profiles compare well with independent
measurements (e.g. Courteau 1996; Courteau \etal 2000).
The relation between $\Vc/\cs$ and $\C28$ is shown in \Fig{VSigC}.  
A fit to the data yields the relation: 

\begin{equation} 
\label{eq:VSC} 
\log(\Vc/\cs)=0.63 - 0.11\C28.
\end{equation} 

The lenticular (brown open circles) and elliptical galaxies (black dots)
in \Fig{VSigC} show signs of a match with the $\Vc-\cs-\C28$ relation 
of spirals, as expected via \Equ{BT87}. 
For spiral disks, the galaxy light concentration correlates 
well with the bulge-to-total (B/T) light ratio\footnote{See 
MacArthur \etal (2003) for details and caveats about the 
parametric and non-parametric computations of $B/T$.}.
For the spiral galaxies in our sample, we find $\Vc/\cs=2(1-B/T)$.  
For pure exponential disks, $B/T \rightarrow 0$, $\C28\sim 2.8$ 
and both expressions for $\Vc/\cs$ based on $B/T$ or $\C28$ 
yield $\Vc/\cs \sim 2$. 


\bigskip 


\section{Discussion}
\label{sec:discuss}

We have demonstrated the dependence of $\Vc/\cs$ on galaxy 
concentration (Eq.~\ref{eq:VSC}), as expected theoretically
(Eq.~\ref{eq:Jeans}).  We now explore the $\Vc-\cs-\C28$ relation 
in the context of dynamical galaxy models. 

\begin{figure*}
\begin{center}
\includegraphics[width=7.0in]{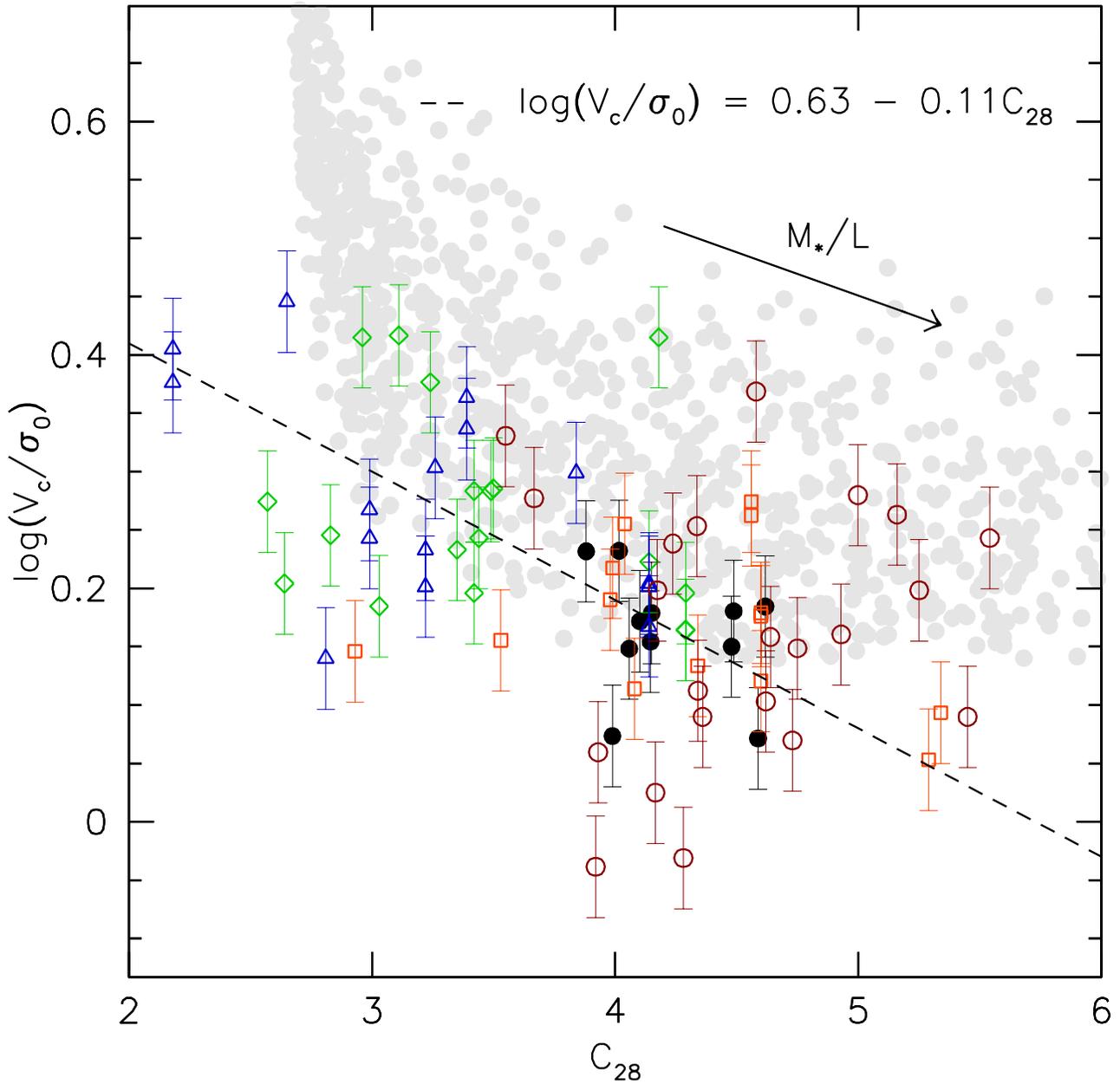}
\caption{Ratio of maximum circular velocity to bulge velocity dispersion 
  versus concentration ratio for 32 spiral, 12 elliptical, and 33 
  lenticular galaxies whose surface brightness profiles and galaxy 
  concentration $\C28$ could be extracted from SDSS images.  Point 
  types are the same as in Fig.~1.  The concentration 
  $\C28=5\log(r_{80}/r_{20})$ uses radii at which 20\% and 80\% 
  of the total galaxy light is contained.
  The light grey filled circles represent self-consistent equilibrium 
  galaxy models by Widrow \& Dubinski (2005) that satisfy the 
  size-luminosity and velocity-luminosity relations of galaxies.
  See \S{4} for the data-model comparison. The dashed-line 
  is a fit through the data for galaxies with measured $\C28$. 
  The solid arrow depicts the trend of {\it stellar} M/L 
  ratio in the $\Vc-\cs-\C28$ plane based on model predictions. }
\label{fig:VSigC}
\end{center}
\end{figure*}

That spheroidal galaxies may trace a similar $\Vc-\cs-\C28$ relation 
as disk galaxies would suggest that: (i) the mass distributions in 
disks and spheroids are self-similar, and (ii) the central velocity 
dispersions are a reflection of similar environments in galaxy's central 
regions such that \Equ{BT87} applies across the full Hubble 
sequence.  Some of the scatter for disk galaxies in \Fig{VSigC}
may be ascribed to disk star contamination in $\cs$'s; however 
we have verified with simulations described below that this effect 
is small ($<5\%$) at least for Milky Way-type galaxies with
pressure-supported bulges. 

We have used the self-consistent equilibrium galaxy models of 
Widrow \& Dubinski (2005 [WD05]) to generate a suite of stable 
galaxy models for a wide range of disk masses and sizes, bulge 
and disk mass-to-light ratios, and dark halo masses. 
The dynamical models, which include an exponential disk 
with anisotropic velocity dispersion, a Hernquist bulge, 
and a NFW dark halo are constructed directly from 
phase-space distribution functions that self-consistently solve 
the collisionless-Boltzmann and Poisson equations.  

A line-of-sight central velocity dispersion and maximum circular 
velocity of the disk are measured to mimic actual observations.  
For each WD05 model, we calculate the maximum circular 
rotation speed, $\Vc$, between $0.5 R_{25}$ and 4 disk scale lengths. 
$R_{25}$ is the radius at which the surface brightness equals $25$ 
mag arcsec$^{-2}$.  We also calculate the line-of-sight velocity 
dispersion $\cs$ within the effective radius of the bulge; 
aperture effects are minimal (just as observed with real data (F06)).  
The galaxy light concentration $\C28$ is calculated by summing up 
all the particles out to the edge of the galaxy and using suitable 
bulge and disk mass-to-light ratios. 

We use a random selection of WD05 models constrained to match the 
size-luminosity and velocity-luminosity relations of galaxies and 
to obey $0.3 < V_{disk}/V_{tot} < 0.85$ (McGaugh 2005; 
Courteau \etal 2006; Dutton \etal 2006). 

The light grey circles in \Fig{VSigC} are the result of 950
dynamically stable WD05 models.  The model boundaries correspond 
to $\C28=2.8$ for pure disks and $\Vc=\sqrt{2}\cs$ 
for isothermal systems.  Galaxies with central cores or strong 
spiral arms, that WD05 simulations do not reproduce, can have 
$\C28<2.8$.  It is intriguing that the measurements for many 
spheroidal (E+S0) galaxies have $\Vc < \sqrt{2} \cs$; this 
departure from theoretical expectations is likely due to
anisotropic velocity ellipsoids and/or to the strength of the central 
cusp.  Also, the models extending to high concentrations likely 
result from the assumption of (highly concentrated) NFW halos.  


Limited as they may be, the WD05 models predict that the distribution 
of {\it stellar} M/L ratios follows roughly the $\Vc-\cs-\C28$ line.
A calibration of this trend awaits stellar M/L
ratios measured from multi-band light profiles or resolved spectra
and accurate stellar population models.

The $\Vc-\cs-\C28$ relation (Eq.~\ref{eq:VSC}) is seemingly a 
product of the size-luminosity and velocity-luminosity relations 
in galaxies and dynamical considerations.  A precise interpretation 
remains, however, beyond the range of our analytical models as a detailed
understanding of galaxy formation and dynamics is still lacking 
(e.g. WD05; Dutton \etal 2006).  For instance, 
dissipational processes (e.g. cooling and feedback, adiabatic 
compression or expansion of the halo) and dynamical effects 
(e.g. angular momentum transfer) that play a crucial role
in setting $\Vc$ and $\cs$ remain poorly constrained. 
As well, the total light concentration is likely a reflection 
of uncertain mass accretion histories.
From an empirical stand-point,
the scatter of scaling relations that involve dynamical parameters, 
(e.g. Faber \& Jackson 1976 or Tully \& Fisher 1977) should clearly 
be reduced by virtue of \Equ{VSC}.  We propose that the $\Vc-\sigma-\C28$ 
relation be used as a new constraint for galaxy formation and dynamical 
models.

It is unfortunate that in 30 years since WKS79, the number of 
galaxies with measured central velocity dispersions and circular 
velocities has only grown from a few dozen to a few hundred.  
The situation is especially dire for elliptical galaxies.
It is imperative for the study of galaxy formation and 
evolution, especially in the wake of large galaxy surveys such 
as SDSS and UKIDSS (Hewett \etal 2006), 
that we remedy this situation with all-sky, deep high-spectral 
resolution ($R>5000$) spectroscopic surveys of thousands 
of galaxies of all Hubble types.  These resolved velocity measurements 
will impact dramatically the study of galaxy scaling relations,
the mass function of galaxies, and tests of baryonic physics to 
map the coupling between central massive objects, bulges, disks,
and halos.

\bigskip We are grateful to James Bullock, Lauren MacArthur, 
and the referee for a careful read and useful comments, and to 
Avishai Dekel, Laura Ferrarese, Ortwin Gerhard, Eric Hayashi, 
Stacy McGaugh, Julio Navarro, and Joanna Woo for stimulating 
discussions.  S.C. and L.M.W. acknowledge the support of NSERC 
through respective Discovery grants.  J.H. was partially supported 
via an NSF AST-0407072 grant.  


\clearpage


\clearpage

\clearpage
\end{document}